\documentclass{midl} % Include author names
\jmlrproceedings{MIDL}{Medical Imaging with Deep Learning}
\jmlrpages{}
\jmlryear{2022}

\jmlrworkshop{Short Paper -- MIDL 2022 submission}
\jmlrvolume{-- Under Review}
\editors{Under Review for MIDL 2022}

\title[PD-UNet CBCT Volumes]{Primal-Dual UNet for Sparse View Cone Beam Computed Tomography Volume Reconstruction}

\midlauthor{\Name{Philipp Ernst\nametag{$^{1,2}$}}\Email{philipp.ernst@ovgu.de}\AND
\Name{Soumick Chatterjee\nametag{$^{1,2}$}}\Email{soumick.chatterjee@ovgu.de}\AND
\Name{Georg Rose\nametag{$^{2,3}$}}\Email{georg.rose@ovgu.de}\AND
\Name{Andreas N\"urnberger\nametag{$^{1}$}}\Email{andreas.nuernberger@ovgu.de}\\
\addr $^{1}$ Faculty of Computer Science, University of Magdeburg, Germany\\
\addr $^{2}$ Research Campus \textit{STIMULATE}, University of Magdeburg, Germany\\
\addr $^{3}$ Institute for Medical Engineering, University of Magdeburg, Germany
}

\begin{document}

\maketitle

\begin{abstract}
In this paper, the Primal-Dual UNet for sparse view CT reconstruction is modified to be applicable to cone beam projections and perform reconstructions of entire volumes instead of slices. Experiments show that the PSNR of the proposed method is increased by 10dB compared to the direct FDK reconstruction and almost 3dB compared to the modified original Primal-Dual Network when using only 23 projections. The presented network is not optimized wrt. memory consumption or hyperparameters but merely serves as a proof of concept and is limited to low resolution projections and volumes.
\end{abstract}

\begin{keywords}
Sparse view CT, cone beam CT, deep learning, convolutional neural networks
\end{keywords}

\section{Introduction}
During CT-guided medical interventions, surgeons and patients are exposed to harmful X-radiation. Keeping the dose low is essential but also results in high noise or streaking artifacts in the reconstructions.
\citet{ernst2022} proposed the Primal-Dual UNet, based on Primal-Dual Network~\cite{adler2018}, for sparse view parallel and fan beam CT reconstruction. In medical interventions, however, surgeons make use of cone beam CT for imaging. Therefore, the main contributions of this work are: (i) modifying the network to process cone beam projections and (ii) reconstruct entire volumes instead of axial slices.

\section{Methods}
The network architecture used in this work is a modified Primal-Dual UNet~\cite{ernst2022}. The two-dimensional convolutions of the dual space blocks were replaced with their three-dimensional counterparts. The two-dimensional UNet in the primal space was replaced with a three-dimensional UNet by replacing convolutions, batch normalizations, average poolings and linear upsamplings with their three-dimensional counterparts. Instead of the parallel or fan beam projection layer, a cone beam geometry (detector: $310\times240$px, $1.232$mm pixel size; SID=$160$mm; SDD=$400$mm) on a circular trajectory was used. The FBP reconstruction layer was replaced with its FDK counterpart.

For comparability, the data normalization, the $L_1$ loss function, the Adam optimizer (lr=1e-3, $\beta_1$=0.9, $\beta_2$=0.999) and the number of epochs (151) were kept the same. The effective batch size was set to 16. Training data was simulated by downsampling LungCT-Diagnosis~\cite{lungct-diagnosis} volumes (42/9/10 for training/validation/test) to cubes with side lengths of $128\mathrm{px}=128\mathrm{mm}$ due to memory limitations. Random flips, rotations and scalings of the volumes were used as augmentation during training. Sparse views were simulated by retaining every 8th or 16th of 360 equiangular projections (called \textit{Sparse 8} or \textit{Sparse 16}, respectively).

\section{Results}
\begin{table}[htbp]
 % The first argument is the label.
 % The caption goes in the second argument, and the table contents
 % go in the third argument.
\floatconts
  {tab:metrics}%
  {\caption{Mean and standard deviation over all axial test slices for \textit{Sparse 16}.}}%
  {\begin{tabular}{lccc}
  \bfseries Model/Method & \bfseries SSIM [\%] & \bfseries PSNR [dB] & \bfseries RMSE [HU]\\
  FDK & 43.54\textpm 8.27 & 17.92\textpm 2.64 & 388.57\textpm 108.15\\
  FDKConvNet & 67.37\textpm 8.99 & 24.72\textpm 1.93 & 177.30\textpm 60.94\\
  Primal-Dual Network & 69.87\textpm7.66 & 25.19\textpm 2.08 & 169.22\textpm 64.35\\
  Primal-Dual UNet & 78.76\textpm 7.50 & 27.93\textpm 2.33 & 128.89\textpm 57.78\\
  \end{tabular}}
\end{table}
\noindent\tableref{tab:metrics} shows the results of the different models evaluated on the test set. All models outperform the direct sparse view FDK reconstruction by a large margin, while the Primal-Dual models further increase the quality compared to FDKConvNet~\cite{jin2017}. The proposed Primal-Dual UNet results in the lowest errors. Wilcoxon signed-rank tests reveal that the proposed model significantly outperforms any other model/method pair-wise (p-value $<0.5\%$).

\begin{figure}[htbp]
 % Caption and label go in the first argument and the figure contents
 % go in the second argument
\floatconts
  {fig:example}
  {\caption{Exemplary axial slice from different models/methods.}}
  {
  \includegraphics[width=\linewidth]{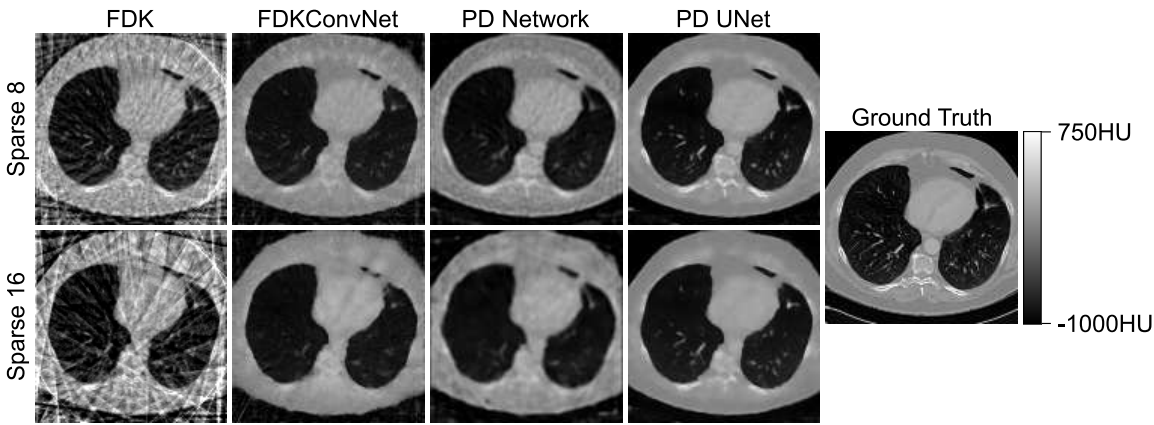}
  }
\end{figure}

\figureref{fig:example} shows an exemplary axial slice from the different models for \textit{Sparse 8} (top row) and \textit{Sparse 16} (bottom row). FDKConvNet does not seem to have learned anatomical structures and merely attempts to suppress streaking artifacts. Primal-Dual Network produces results that look blurrier with more low frequency noise than FDKConvNet's outputs but anatomical structures, e.g. costal cartilage, are preserved better. The reconstructions of Primal-Dual UNet are superior compared to Primal-Dual Network. Tissues with high attenuation coefficients are clearly distinguishable from soft tissues and edges are well preserved, e.g. vertebrae, even for the higher sparsity factor \textit{Sparse 16}.

\section{Discussion and Conclusion}
The proposed Primal-Dual UNet for cone beam reconstruction not only outperforms other methods -- Primal-Dual Network in particular -- in quality but also in memory requirements and is more than twice as fast during both training and inference while retaining data consistency wrt. the cone beam projections, as opposed to FDKConvNet. Moreover, the training of the proposed network is much more stable compared to Primal-Dual Network.
% Reduction of X-ray exposure during CT-guided medical interventions is essential for surgeons and patients not to develop harmful diseases, but results in lower quality reconstructions compared to using full dose. 
However, the main limitation is still the memory consumption: with enabled mixed precision, the inference takes $\sim$9GB of GPU RAM for even these unrealistically low resolution volumes and projections and a batch size of 1. Training consumes even more space: a \textit{Sparse 4} version of Primal-Dual Network did not even fit into the 48GB of an Nvidia RTX A6000.

Since usually, not the entire volume needs to be reconstructed during an intervention, future work will focus on reducing the memory requirements by only reconstructing volumes of interest. Moreover, this preliminary work is based on simulations and has to be evaluated for real cone beam CT data.
The Pytorch implementation is available on Github\footnote{\url{https://github.com/phernst/pd-unet-conebeam}}.

% Acknowledgments---Will not appear in anonymized version
\midlacknowledgments{This work was supported by the ESF (project no. ZS/2016/08/80646).}

\bibliography{midl-shortpaper}

\end{document}